\documentclass[twocolumn]{aastex631}
\usepackage[utf8]{inputenc}
\DeclareUnicodeCharacter{2212}{-}

\usepackage{hyperref}

\received{2024 May 22}
\revised{2024 July 1}
\accepted{2024 July 9}
\published{2024 August 2}
\shortauthors{Chen et al.}
\begin{document}

\title{Early-phase simultaneous multiband observations of the Type II supernova SN~2024ggi with Mephisto}

\author[0009-0000-4068-1320]{Xinlei Chen}
\affiliation{South-Western Institute for Astronomy Research, Yunnan University, Kunming, Yunnan 650504, People's Republic of China}

\author[0000-0001-7225-2475]{Brajesh Kumar}
\affiliation{South-Western Institute for Astronomy Research, Yunnan University, Kunming, Yunnan 650504, People's Republic of China}

\author[0000-0002-8700-3671]{Xinzhong Er}
\affiliation{South-Western Institute for Astronomy Research, Yunnan University, Kunming, Yunnan 650504, People's Republic of China}

\author[0000-0001-5737-6445]{Helong Guo}
\affiliation{South-Western Institute for Astronomy Research, Yunnan University, Kunming, Yunnan 650504, People's Republic of China}

\author[0000-0001-6374-8313]{Yuan-Pei Yang}
\affiliation{South-Western Institute for Astronomy Research, Yunnan University, Kunming, Yunnan 650504, People's Republic of China}

\author[0000-0003-2240-7031]{Weikang Lin}
\affiliation{South-Western Institute for Astronomy Research, Yunnan University, Kunming, Yunnan 650504, People's Republic of China}

\author[0009-0006-1010-1325]{Yuan Fang}
\affiliation{South-Western Institute for Astronomy Research, Yunnan University, Kunming, Yunnan 650504, People's Republic of China}

\author[0000-0002-8109-7152]{Guowang Du}
\affiliation{South-Western Institute for Astronomy Research, Yunnan University, Kunming, Yunnan 650504, People's Republic of China}

\author[0000-0001-5561-2010]{Chenxu Liu}
\affiliation{South-Western Institute for Astronomy Research, Yunnan University, Kunming, Yunnan 650504, People's Republic of China}

\author{Jiewei Zhao}
\affiliation{Department of Astronomy, Yunnan University, Kunming, Yunnan 650504, People's Republic of China}

\author{Tianyu Zhang}
\affiliation{South-Western Institute for Astronomy Research, Yunnan University, Kunming, Yunnan 650504, People's Republic of China}

\author{Yuxi Bao}
\affiliation{South-Western Institute for Astronomy Research, Yunnan University, Kunming, Yunnan 650504, People's Republic of China}

\author[0009-0006-5847-9271]{Xingzhu Zou}
\affiliation{South-Western Institute for Astronomy Research, Yunnan University, Kunming, Yunnan 650504, People's Republic of China}

\author[0009-0002-7625-2653]{Yu Pan}
\affiliation{South-Western Institute for Astronomy Research, Yunnan University, Kunming, Yunnan 650504, People's Republic of China}

\author[0009-0001-5288-1108]{Yu Wang}
\affiliation{South-Western Institute for Astronomy Research, Yunnan University, Kunming, Yunnan 650504, People's Republic of China}

\author[0009-0003-6936-7548]{Xufeng Zhu}
\affiliation{South-Western Institute for Astronomy Research, Yunnan University, Kunming, Yunnan 650504, People's Republic of China}

\author[0000-0002-6252-3750]{Kaushik Chatterjee}
\affiliation{South-Western Institute for Astronomy Research, Yunnan University, Kunming, Yunnan 650504, People's Republic of China}

\author[0000-0003-0394-1298]{Xiangkun Liu}
\affiliation{South-Western Institute for Astronomy Research, Yunnan University, Kunming, Yunnan 650504, People's Republic of China}

\author[0000-0002-0409-5719]{Dezi Liu}
\affiliation{South-Western Institute for Astronomy Research, Yunnan University, Kunming, Yunnan 650504, People's Republic of China}

\author[0000-0003-1713-0082]{Edoardo P. Lagioia}
\affiliation{South-Western Institute for Astronomy Research, Yunnan University, Kunming, Yunnan 650504, People's Republic of China}

\author[0000-0002-6373-770X]{Geeta Rangwal}
\affiliation{South-Western Institute for Astronomy Research, Yunnan University, Kunming, Yunnan 650504, People's Republic of China}

\author{Shiyan Zhong} 
\affiliation{South-Western Institute for Astronomy Research, Yunnan University, Kunming, Yunnan 650504, People's Republic of China}

\author[0000-0002-2510-6931]{Jinghua Zhang}
\affiliation{South-Western Institute for Astronomy Research, Yunnan University, Kunming, Yunnan 650504, People's Republic of China}

\author[0000-0001-5258-1466]{Jianhui Lian}
\affiliation{South-Western Institute for Astronomy Research, Yunnan University, Kunming, Yunnan 650504, People's Republic of China}

\author[0000-0002-7714-493X]{Yongzhi Cai} 
\affiliation{Yunnan Observatories, Chinese Academy of Sciences, Kunming 650216, People's Republic of China}
\affiliation{Key Laboratory for the Structure and Evolution of Celestial Objects, Chinese Academy of Sciences, Kunming 650216, People's Republic of China}
\affiliation{International Centre of Supernovae, Yunnan Key Laboratory, Kunming 650216, People's Republic of China}

\author{Yangwei Zhang}
\affiliation{South-Western Institute for Astronomy Research, Yunnan University, Kunming, Yunnan 650504, People's Republic of China}

\author[0000-0003-1295-2909]{Xiaowei Liu}
\affiliation{South-Western Institute for Astronomy Research, Yunnan University, Kunming, Yunnan 650504, People's Republic of China}

\correspondingauthor{Brajesh Kumar}
\email{brajesh@ynu.edu.cn, brajesharies@gmail.com} 
\email{xer@ynu.edu.cn}
\email{x.liu@ynu.edu.cn}


\begin{abstract}
We present early-phase good-cadence (hour-to-day) simultaneous multiband ($ugi$ and $vrz$ bands) imaging of the nearby supernova SN~2024ggi, which exploded in the nearby galaxy, NGC~3621. A quick follow-up was conducted within less than a day after the explosion and continued $\sim$23 days. The $uvg$-band light curves display a rapid rise ($\sim$1.4 mag day$^{-1}$) to maximum in $\sim$4 days and absolute magnitude $M_{g}\sim$--17.75 mag. The post-peak decay rate in redder bands is $\sim$0.01 mag day$^{-1}$. Different colors (e.g., $u-g$ and $v-r$) of SN~2024ggi are slightly redder than SN~2023ixf. A significant rise ($\sim$12.5 kK) in black-body temperature (optical) was noticed within $\sim$2 days after the explosion, which successively decreased, indicating shock break out inside a dense circumstellar medium (CSM) surrounding the progenitor. Using semianalytical modeling, the ejecta mass and progenitor radius were estimated as 1.2 $M_\sun$ and $\sim$550 $R_\sun$. The archival deep images ($g,r,i and z$ bands) from the Dark Energy Camera Legacy Survey were examined, and a possible progenitor was detected in each band ($\sim$22--22.5 mag) and had a mass range of 14--17 $M_\sun$.
\end{abstract}

\section{Introduction}\label{intro}

In recent years, the early detection of several nearby Type II supernovae (SNe) shortly after their explosions has drawn substantial attention from the time-domain community. These events belong to the core-collapse supernovae (CCSNe) group. The spectro-photometric properties lead to further subclassification of Type II events (e.g., IIP, IIL, IIb, and IIn), but the presence of Balmer lines in their spectra is a common feature \citep{1997filippenko}. Their light-curve evolution is mainly governed by the hydrogen envelope mass retained before the explosion and the explosion energy \citep{Dessart-2013}. Based on direct identifications, it is believed that Type IIP/IIL SNe originate from the red supergiant (RSG) progenitors (zero-age main-sequence mass range $\gtrsim$8--25 M$_\odot$) at the end stage of their evolution \citep{Smartt2009, Smartt2015}. The earliest postexplosion observation (minutes--hours to days) is crucial in these events as it retains information about the shock-breakout (SBO)/shock-cooling emission and the preexisting circumstellar medium (CSM) surrounding the progenitor \citep[see][]{Waxman2017hsn..book, 2018moriya, 2018ApJ...858...15M}.

The presence or absence of CSM strongly influences the appearance of light curves and spectral features, especially during the early epochs \citep{2011chevalier}. SBO is revealed in the form of ultraviolet (UV)/X-ray radiation immediately after the gravitational collapse of the stellar core, lasting for minutes to several hours. The propagating shock finally breaks out of the stellar surface and ionizes the surrounding CSM, which consequently appears as ``ﬂash" features in the spectra \citep{Quimby-2007ApJ...666.1093Q, 2014galyam, 2017yaron, Jacobson-2024arXiv240302382J}. Subsequently, shock-cooling emission occurs in UV/optical wavelengths, where the radiation is driven by photons escaping from deeper layers in the expanding envelope. The duration of shock-cooling may survive several days, where the cooling rate depends on the stellar radius and the composition of the envelope \citep{Ganot-2016ApJ, Morag-2023}. Due to the short timescale, the early-phase observations of Type II are very limited. Only a handful of events have been adequately monitored and studied in detail, e.g., SN~2013fs \citep{2017yaron, 2018MNRAS.476.1497Bullivant}, SN~2016gkg \citep{Tartaglia-2017ApJ...836L..12T}, SN~2018zd \citep{Hiramatsu-2021NatAs...5..903H, Jujia-2020MNRAS.498...84Z}, SN~2023ixf \citep{2023galan, 2023hiramatsu, 2023hosseinzadeh, Smith-2023ApJ, 2023teja, 2023vasylev, Bostroem-2023ApJ, Yamanaka-2023PASJ, Zhang-2023SciBu..68.2548Z, 2023zimmerman, Li-2024Natur.627..754L, Avinash-2024, Yuan_Pei-sn23ixf}. Investigating multiband early-phase light curves with a good cadence (hour to day) is important to determine various parameters of the progenitor \citep[see][and references therein]{Rabinak-2011, 2023hosseinzadeh}.

\begin{figure}
\centering
\includegraphics[width=8.2cm]{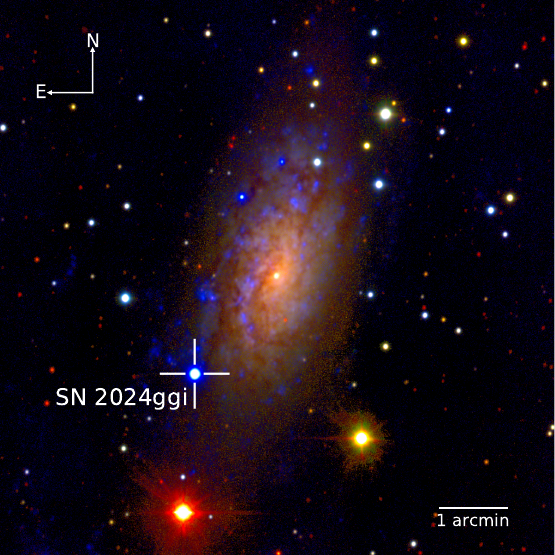}
\caption{SN~2024ggi in the spiral galaxy NGC~3621 is marked. This color composite image is created using several night stacked frames of the Mephisto three bands ($u,g,i$). Colors are adjusted to enhance the features. The SN is indicated with a cross-hair, and the displayed image size is $\sim$8\,\arcmin $\times$ 8\arcmin.}
\label{fig:picture}
\end{figure}

SN~2024ggi (ATLAS24fsk, GOTO24aig,  BGEM J111822.10-325015.1, PS24brj) is another nearby event after last year's SN~2023ixf \citep{Itagaki-2023TNS}, discovered in the early phase after the explosion and it provides an excellent opportunity to investigate the observational properties comprehensively, including its progenitor.
In this Letter, we present early-phase multiband simultaneous imaging of SN~2024ggi. The SN was discovered \citep{Tonry-2024, Srivastav-2024} in NGC~3621 on MJD 60411.14 (UT 2024 April 11.14) by the ATLAS survey group with a discovery magnitude of 18.915 (orange-ATLAS). The fast rise in SN magnitudes was confirmed in further monitoring \citep{Chen-2024, Kumar-2024TNS}. The early spectra taken within a few hours suggest a Type II SN with flash features and the redshift $z_s=0.002435$ \citep{Hoogendam-2024TNS, Zhai-2024TNS}. Using the Gravitational-wave Optical Transient Observer \citep[GOTO;][]{Steeghs-2022} imaging, \citet{Killestein-2024} put a strong constraint on the explosion epoch MJD~60410.80, which is consistent as estimated by \citet{Jacobson-2024ggi}. We have adopted it further in this Letter.

In recent studies, several authors investigated the early-phase multiwavelength observations of SN~2024ggi \citep{Chen-2024C, Jacobson-2024ggi, Pessi-sn24ggi, Shrestha-2024S, Jujia-2024Z}. The initial spectra (within a few hours after the explosion) exhibit prominent narrow emission lines (H\,{\sc i}, He\,{\sc i}, C\,{\sc iii}, and N\,{\sc iii}), indicating the photoionization of dense, optically thick CSM. In the successive spectra ($>$1 day), the highly ionized spectral lines such as N\,{\sc iv}, N\,{\sc v}, C\,{\sc iv}, O\,{\sc iv}, O\,{\sc v} emerged \citep[see][]{Jacobson-2024ggi, Pessi-sn24ggi, Shrestha-2024S, Jujia-2024Z}. X-ray emission was also detected from SN~2024ggi in a few days after the discovery by different X-ray missions \citep{Margutti-2024ATel, Lutovinov-2024ATel, Zhang2024ATel}, which suggests SN shock interaction with the dense medium.

In Section~\ref{sec:obs}, we provide the observation and data analysis of the Multi-channel Photometric Survey Telescope (Mephisto) follow-up. Section~\ref{result} discusses the light-curve properties, including the color and bolometric light-curve evolution. The progenitor properties and summary are presented in Section~\ref{SN-progenitor} and Section~\ref{summary}, respectively.

\section{Mephisto observations and data reduction}\label{sec:obs}

The photometric observations of SN~2024ggi were initiated with the 1.6 m Mephisto (X. Liu et al. 2024, in preparation) immediately after receiving the Transient Name Server notice.
However, due to limitations of the southern sky target and weather, the first frame was obtained at MJD\,=\,60411.64 (UT 2024 April 11.64). After detecting SN~2024ggi in our initial images, we continued monitoring the target until its availability in the sky. The SN was followed up on consecutive nights whenever the conditions permitted. A stacked image of Mephisto frames marking the SN location is shown in Fig.~\ref{fig:picture}. 

It is important to highlight that Mephisto is a wide-field multi-channel telescope, the first of its type worldwide. The facility is located at Lijiang Observatory (IAU code: 044) of Yunnan Astronomical Observatories, Chinese Academy of Sciences (CAS), with longitude $100^{\circ}01'48''$ east, latitude $26^{\circ}41'42''$ north and altitude 3200m \citep[more details about the site can be found in][]{Wang-2019RAA}. The South-Western Institute for Astronomy Research, Yunnan University, operates the Mephisto facility, which is presently in the commissioning phase. However, it is already contributing with good-quality scientific data \citep[see, e.g.,][]{Chen-2024, Yuan_Pei-sn23ixf}. The facility is equipped with three-channel CCD cameras, viz., blue ($uv$), yellow ($gr$), and red ($iz$) channels, and is capable of performing simultaneous observations in $ugi$ or $vrz$ optical bands at each pointing. The wavelength coverage of the $u,v,g,r,i$ and $z$ filters are $320-365$, $365-405$, $480-580$, $580-680$, $775-900$, and $900-1050$ nm, respectively (see Appendix).

Since the location of SN~2024ggi is in the Southern Hemisphere (R.A.\,=\,11$^{\rm h}$~18$^{\rm m}$~22\fs09 and Dec.\,=\,--32\degr ~50\arcmin ~15\farcs29, J2000), there is a limitation for Mephisto to observe the source whole night and only about one-hour observation could be performed. We allocated all the possible observing time for SN~2024ggi during the first week after its discovery. The two sets of filters, $ugi$ and $vrz$, were executed consecutively.

The raw frames were preprocessed (bias subtraction, flat fielding, and cosmic ray removal) using a preprocessing pipeline developed for the Mephisto (Y. Fang et al. in preparation). Point spread function (PSF) photometry was performed on the cleaned images to get the instrumental magnitudes. Gaia BP/RP low-resolution spectra (XP spectra) are obtained by the Blue and Red Photometers on the Gaia satellite \citep{2023A&A...674A...2D, 2023A&A...674A...3M}. We utilized the Gaia XP spectra of nonvariable stars to perform photometric calibration. Because Gaia XP spectra exhibit systematic errors that depend on magnitude, color, and extinction, particularly at wavelengths below 400 nm, we used the corrected Gaia XP spectra from \citet{2024ApJS..271...13H}. The wavelength coverage of corrected Gaia XP spectra is from 336 to 1020 nm, which does not fully cover our $u$ and $z$ bands. Thus, we extrapolated the spectra for the $u$ ($336-364$ nm) and $z$ bands. We computed each band's synthetic magnitude in the AB system by convolving the spectra with the transmission efficiency. We calculated the difference $\Delta$m between the instrumental and synthetic magnitudes for the nonvariable stars and used the median $\Delta$m to calibrate our photometric measurements. Here, we emphasize that the magnitude errors mainly include the photon noise and detector noise. The uncertainties in the photometric calibration and the magnitude zero points are usually $\sim$0.01 mag and less than 0.03 mag in $vgri$ and $uz$ bands, respectively. The possible flux contamination due to the host galaxy has not been removed as the SN is very bright, and we consider that host flux has a minimal effect at this stage.  

The foreground extinction from the Milky Way in the line of sight of SN~2024ggi is $E(B-V)_{\rm MW}\,=\,$0.070 mag \citep{Schlafly11}. For the extinction from the host galaxy, \citet{Jacobson-2024ggi} obtained $E(B-V)_{\rm host}=0.084\pm 0.018$ mag by calculating the equivalent widths (EWs) of Na\,{\sc i} D2 and D1. According to the total extinction of $E(B-V)_{\rm total}={0.154}\,\pm\,0.018$ mag and the extinction law of \citet{Fitzpatrick1999PASP} ($R_V$=3.1), we finally obtained the extinction values of 0.762, 0.695, 0.496, 0.390, 0.248, and 0.202 mag for the $u,v,g,r,i,z$-bands, respectively. An estimate for the distance of the host galaxy NGC~3621 over different methods (Cepheids, TRGB, and Tully-Fisher) is adopted \citep{2013AJ....146...86T}, which is $6.73$ Mpc. The distance modulus is $29.14$. 

\begin{figure}
\centerline{\includegraphics[width=8cm]{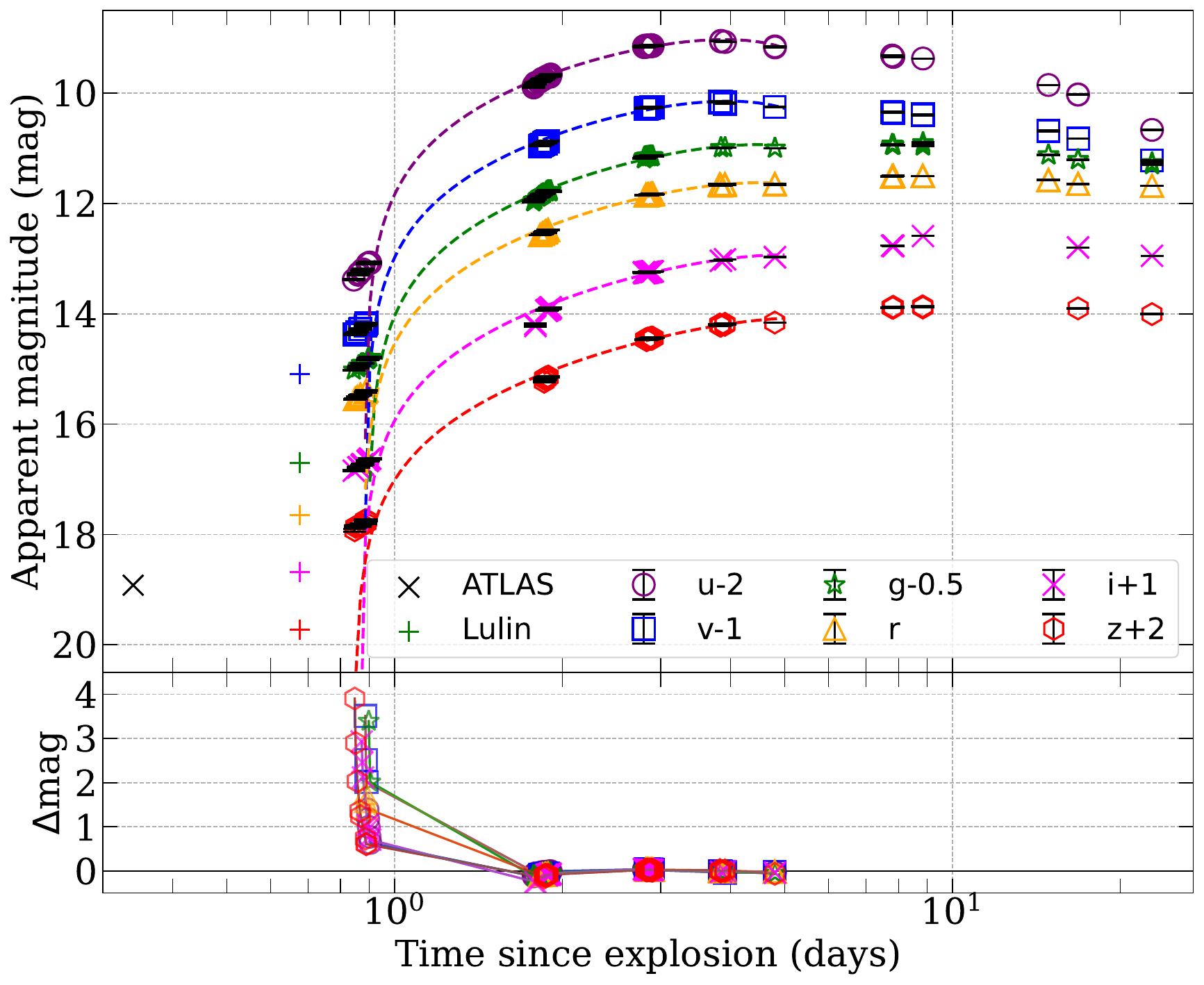}}
\caption{Top panel: $uvgriz$-band light-curve evolution of SN~2024ggi. Indicated offsets have been applied to the apparent magnitudes for clarity. The cross marks the time and magnitude of the discovery by ATLAS \citep{Tonry-2024, Srivastav-2024} and the plus symbols indicate the early observations by Lulin observatory \citep{Chen-2024}. One should notice that the filter systems used by ATLAS and Lulin differ from those used by Mephisto. The dashed lines are the fitting light curves using a single power-law model (see Section~\ref{LC-properties}). Bottom panel: the residuals between the observed and fitted light curves in six bands by Mephisto.}
(The data used to create this ﬁgure are available in the online article.)
\label{fig:fig_lc}
\end{figure}

\begin{figure}
\includegraphics[width=8.5cm]{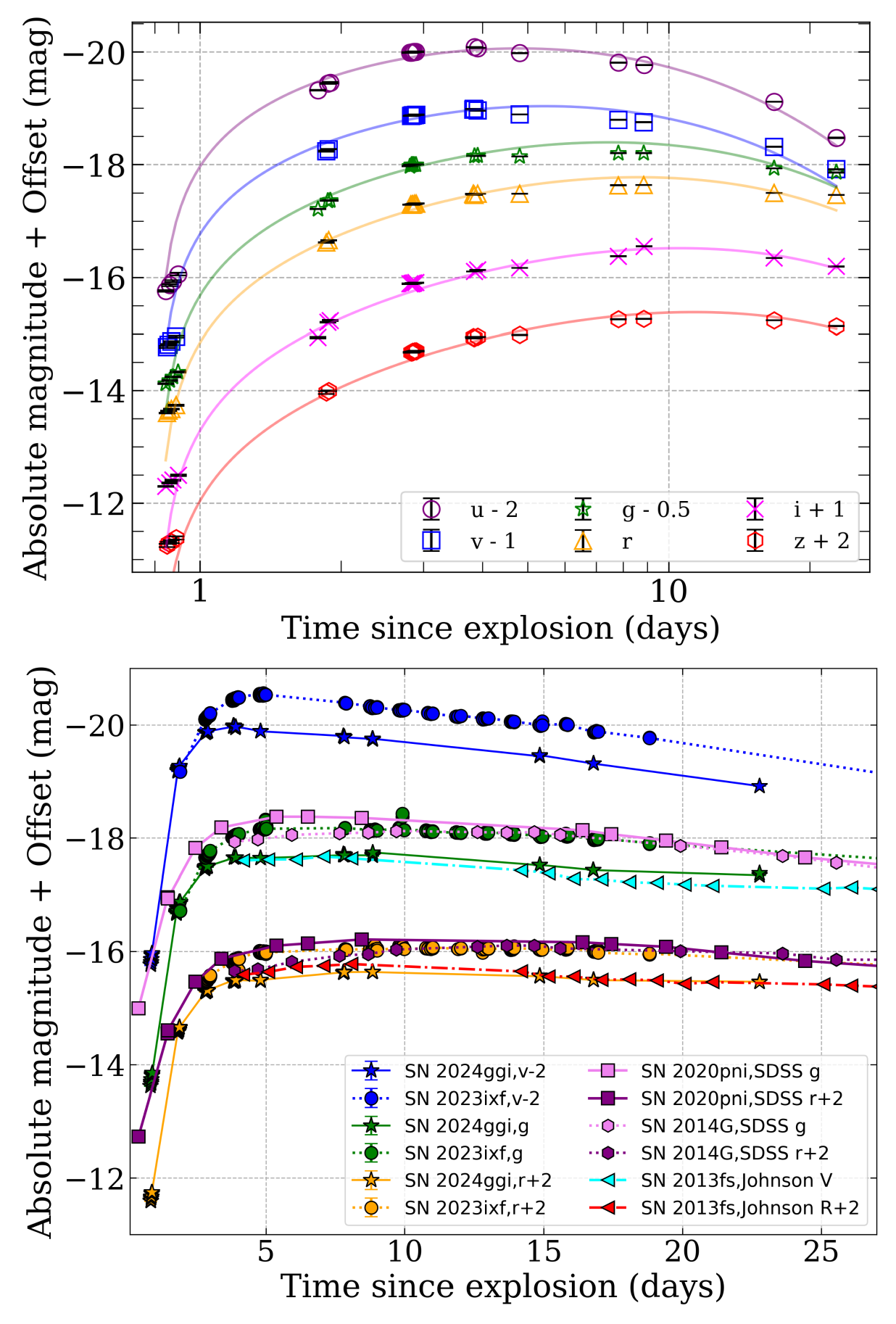}
\caption{Top panel: fitting the shock-cooling model to the multiband light curves of SN~2024ggi (see Section~\ref{LC-properties}). Bottom panel: the multiple-band light curves of SN~2024ggi (stars and solid lines) and other Type II SNe. The blue, green, and orange points show observations with Mephisto filters ($v,g,r$).
SN~2023ixf: circles and dotted lines; SN~2020pni: squares and solid lines; SN~2014G: pentagons and dotted lines; SN~2013fs: triangles and dotted-dashed lines. The other color points show observations with filters of SDSS $g,r$ (violet and purple) or Johnson $V,R$ (cyan and red).} 
\label{fig:shock_cooling}
\end{figure}

\section{Results}\label{result}
\subsection{Light-curve properties}\label{LC-properties}

In this section, we examine the light-curve properties of SN~2024ggi, compare them with other similar types of events, and also estimate different parameters by modeling the light curves. In Fig.\,\ref{fig:fig_lc}, the $uvgriz$-band light curves of SN~2024ggi observed by Mephisto are displayed. It is interesting to note that the $u$ and $v$-band reached a peak within 4 days and then started to decrease. We used a fireball model \citep{Goldhaber-1998, Riess1999AJ....118.2668R} to fit the initial five days of our photometric data points (dashed lines in the figure) to understand the light-curve evolution at the early phase. It should be noted that the black-body temperature decreases rapidly after 2 days; however, in this fitting, we assume that the temperature is constant and the ejecta are in homologous expansion. Although the first assumption is not that certain at this early phase, it is a usual consideration in SNe explosion time estimation \citep[e.g.,][]{Riess1999AJ....118.2668R}. The flux is expressed as $f\propto(t-t_0)^2$ \citep{2023hosseinzadeh}. The fitting curves can represent the observed light curves from the second to the fifth day. However, such a fitting yields the first light epoch as MJD~60411.61, which is about 11 hours later than the discovery epoch reported by ATLAS and GOTO observations (see Section~\ref{intro}).

The residual magnitudes for the first five days from the fitting curves are shown in the bottom panel of Fig.\,\ref{fig:fig_lc}. Even with our observations from the first day, a relatively large inconsistency can be seen in all six bands. During the first day, the differences between the observed and fitted $v$ and $z$--band light curves reached 4 magnitudes. It is clear that a single power law could not describe the light-curve in any band in the early phase. The disagreement between the observed and fitting first light epoch was also noticed in the case of SN~2023ixf \citep[cf. $\sim$11 hours,][]{2023hosseinzadeh}, and several explanations were provided, such as the rising light-curve is due to an eruption or instability activity in the progenitor before the gravitational collapse of the core (mass loss) and/or strong influence of CSM interaction. 
To constrain the progenitor properties and explosion parameters of SN~2024ggi, the shock-cooling model of \citet{Morag-2023} was fitted to our early photometry using a Markov Chain Monte Carlo routine implemented in the Light Curve Fitting Package \citep{2023zndo...8049154H}. The estimated parameters are listed in Table~\ref{Fit-para}, and the fitted models are displayed in Fig.~\ref{fig:shock_cooling} (top panel).
The best-fit progenitor radius of SN~2024ggi is $547^{+14}_{-10}$ $R_\sun$, which is larger than the progenitor radius of SN~2023ixf \citep[$410\pm10$ $R_\sun$,][]{2023hosseinzadeh} but consistent with $517^{+13}_{-19}$ $R_\sun$ estimated for SN~2024ggi by \citet{Shrestha-2024S}. 

\begin{table}
\centering
\caption{The best fitting values of the shock cooling model (see Section~\ref{LC-properties}).} \label{Fit-para}
\begin{tabular}{c|c|c} \hline
Parameter      &Variable   &Result   \\ \hline
Shock speed ($10^{8}$cm/s)    &$v_s$ &$5.60^{+0.09}_{-0.13}$   \\
Envelope mass (M$_\odot$) &$M_{env}$  &$1.66^{+0.03}_{-0.03}$   \\
Ejecta mass $\times$ factor (M$_\odot$) &$f_\rho M$ &$1.2^{+0.3}_{-0.2}$    \\
Progenitor radius (R$_\sun$) &$R$     &$547^{+14}_{-10}$ \\
Explosion epoch (day) &$t_0$    &$60411.61^{+0.0001}_{-0.0001}$ \\ \hline
\end{tabular}\\
Note: Here, $f_\rho M$ is a product of ejecta mass ($M$) and a numerical factor $f_\rho$, which is of the order of unity. $f_\rho M$ appears together in the shock-cooling formalism and is considered a single parameter in the model \citep[see][]{2023hosseinzadeh}.
\end{table}

\begin{figure}
\includegraphics[width=8.4cm]{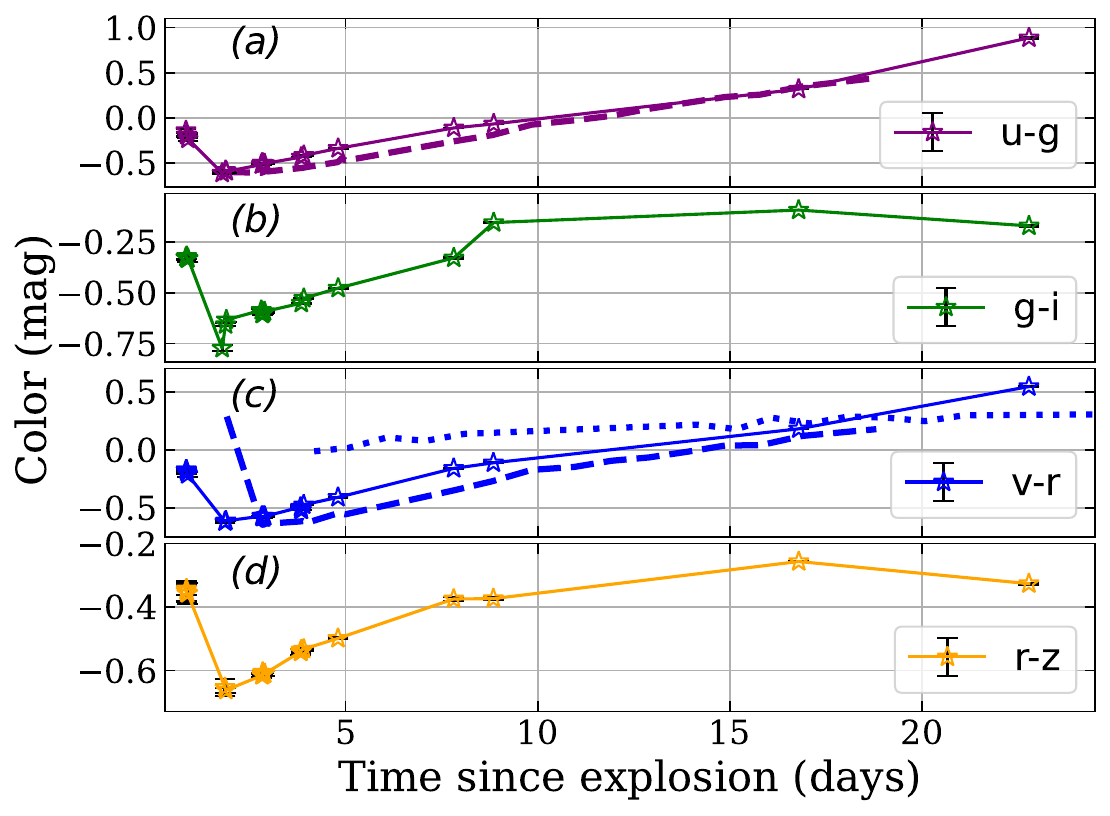}
\caption{The color evolution of SN~2024ggi is shown with star symbols. For comparison, the color curves of two similar type of SNe are over-plotted: SN~2023ixf obtained from Mephisto (dashed) and SN~2013fs (dotted). The filters of SN~2013fs are Johnson $V$ and $R$.}
\label{fig:colorcurve}
\end{figure}

In Fig.~\ref{fig:shock_cooling} (bottom panel), we compare the multiple band light curves of SN~2024ggi with other well-studied events. Here, only those events are considered that were discovered early with flash features, such as SNe~2023ixf \citep{Yuan_Pei-sn23ixf}, 2020pni \citep{2022terreran}, 2014G \citep{2016terreran} and 2013fs \citep{2017yaron}. It can be clearly seen that SN~2024ggi evolved rapidly and reached maximum light within 4 days. The peak absolute magnitude is $-17.75\pm 0.2$ in the $g$--band, while SN~2023ixf reached the peak after 5 days ($\sim$$M_{g}$=--18.43). Both are toward the brighter end of the average peak absolute magnitude ($V$-band) of normal Type II SNe \citep[--16.74 mag,][]{2014anderson} and \citep[--16.89 mag,][]{Galbany-2016AJ....151...33G}. It is also notable that the rise time (in the $u$ and $v$ bands) of $\sim$4 days in SN~2024ggi is significantly shorter than the average peak rise time of $\sim$10 days ($V$-band) of normal Type II events \citep{2016valenti}. The $r,i$ and $z$ light curves rise to the maximum 8.5\,$\pm$\,0.5 days after the explosion.

\subsection{Color and Pseudo-bolometric Light-curve evolution}\label{color}

The color evolution of SNe provides substantial information about the temperature, chemical composition of the ejecta, and shock interaction with CSM in the vicinity of the progenitor. Twofold color evolution has been observed in normal Type II events. As the surface temperature decreases, the color indices gradually approach the redder side in the first few weeks, which later become shallower \citep{Galbany-2016AJ....151...33G}. In large sample studies, a moderate correlation has been observed between the different phase color indices and the explosion energy and $^{56}$Ni mixing \citep{Martinez-2022}. The early-phase multicolor information of CSM-interacting SNe is comparatively limited in the literature. The CSM interacting events display a reversal in UV--optical colors from red--blue to blue--red in about 5 days after the explosion \citep{Jujia-2020MNRAS.498...84Z, Hiramatsu-2021NatAs...5..903H}. 

Taking advantage of simultaneous Mephisto imaging ($ugi$ and $vrz$), we investigated the four colors ($u-g, g-i, v-r$ and $r-z$) of SN~2024ggi. The intrinsic colors are displayed in Fig.\,\ref{fig:colorcurve}. It is evident that the SN became bluer rapidly in all colors during the first two days. The color indices $u-g, g-i, v-r$, and $r-z$ changed from $\sim$--0.4, --0.3, --0.1, and --0.35 mag to $\sim$--0.6, --0.75, --0.6, and --0.7 mag in this period, respectively. Here, we note that the color accuracy is approximately 3 \% after considering error propagation of photometric measurements (see Section~\ref{sec:obs}). The colors of the $u-g$ and $v-r$ became redder after 2 days and continued until the observations. However, there is a reversal in the evolution of $g-i$ and $r-z$ colors. In panel (c) of Fig.\,\ref{fig:colorcurve}, the $v-r$ color of SNe 2013fs, 2023ixf (where flash features were identified) is overplotted for a comparison. It indicates that the color evolution rate in SN~2023ixf is faster than other events in the beginning, and both SN~2024ggi and SN~2023ixf are bluer than other SNe. After $\sim$2.5 days, all colors gradually became redder. SN~2024ggi has a higher excess in all color indices. The initial blue colors signify high temperature and flash-ionization spectral lines \citep{Jacobson-2024ggi, Pessi-sn24ggi, Shrestha-2024S, Jujia-2024Z}. 

\begin{figure}
\includegraphics[width=8.5cm]{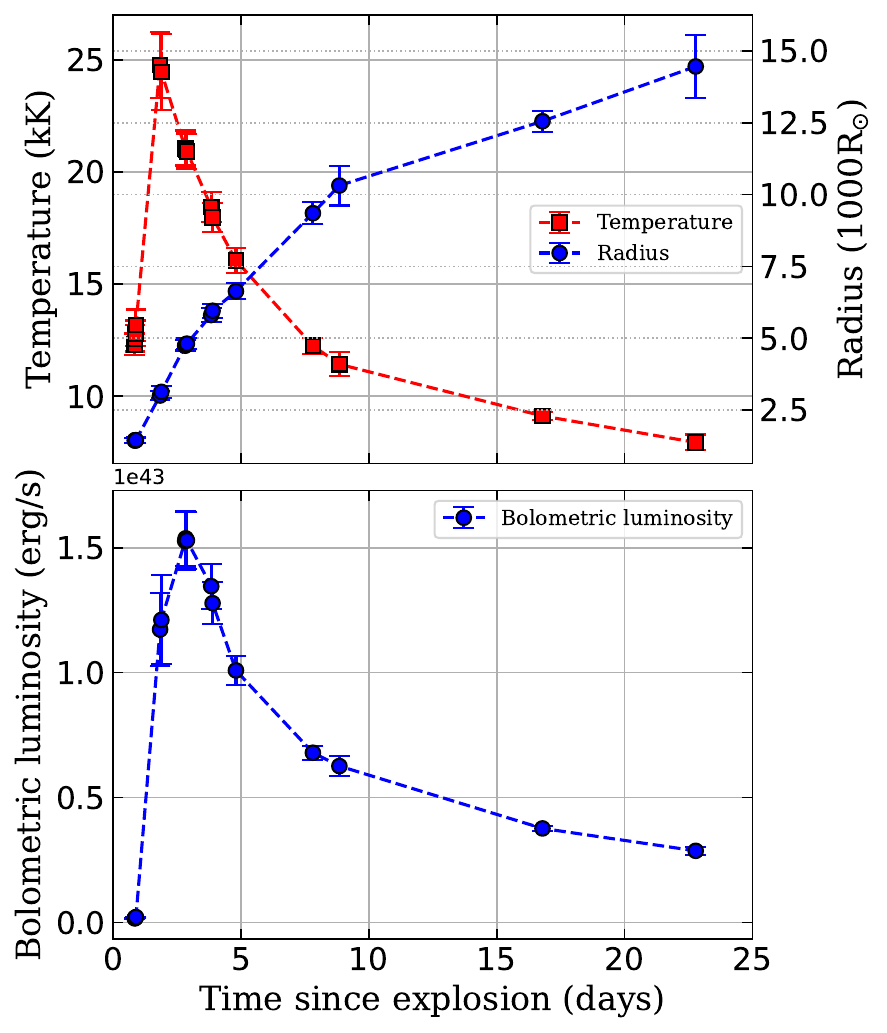}
\caption{Top panel: the evolution of BB temperature and radius of SN~2024ggi. Bottom panel: the bolometric luminosity.}
\label{fig:temp&radius}
\end{figure}

The pseudobolometric light-curve of SN~2024ggi was constructed using SuperBol \citep{Nicholl2018}. The extinction-corrected $uvgriz$ magnitudes and distance 6.73 Mpc were used as input parameters. Here, it is emphasized that the UV/NIR contributions were not included (due to unavailability of data in these wavelengths) in deriving the bolometric magnitudes, but they may have reasonable contributions in some cases like SN~2023ixf \citep{Jacobson-2024ggi, 2023teja, 2023zimmerman, Yamanaka-2023PASJ, Avinash-2024}. The pseudobolometric light-curve is plotted in Fig.~\ref{fig:temp&radius} (bottom panel), which reached to maximum on day 2.6 $\pm$ 0.2 with bolometric luminosity of 1.54 $\pm$ 0.12 $\times$ 10$^{43}$ erg s$^{-1}$. The temperature and radius parameters that resulted in the black body fit are shown in Fig.~\ref{fig:temp&radius} (upper panel). It is worth noting that the temperature evolution, in the beginning, does not follow the usual trend of normal Type II SNe where temperature cools rapidly with increasing photospheric radius after the shock break out \citep{1977falk}. In a day, the temperature reached from 12500 K to 25000 K. A similar rise was noticed by \citet{Jacobson-2024ggi, Chen-2024C, Shrestha-2024S} in SN~2024ggi, and also in SN~2023ixf \citep{2023zimmerman, Avinash-2024} (although UV flux was included). 
Such features of temperature evolution in SN~2024ggi are indicative of a possible shock break out inside a compact and dense CSM surrounding the progenitor \citep[see][]{2018forster, 2018moriya, Hiramatsu-2021NatAs...5..903H, 2023hiramatsu, Li-2024Natur.627..754L, Jujia-2020MNRAS.498...84Z}.

\section{Possible progenitor of SN~2024ggi}\label{SN-progenitor}

In recent years, the direct detection of SNe II-P progenitors has been possible in a few dozen cases thanks to high-resolution images from space and ground-based facilities \citep[see, e.g.,][]{Smartt2009, Smartt2015, Tartaglia-2017ApJ...836L..12T, 2017vandyk}. Several studies of the SN~2023ixf progenitor were made due to its proximity to the Earth \citep{Kilpatrick-2023ApJ, Chenxu-2023ApJ, 2023dong, 2023ApJ...952L..30J, 2023niu, Pledger-2023ApJ, 2023ATel16042....1S, 2023panjkov, Soraisam-2023S, 2024bersten, Neustadt-2024N, Ransome-2024R, 2023vandyk}. As the RSG progenitors have a large mass range \citep{2017PASA...34...58E, Davies-2020D}, any new possible identification is important to better constrain the mass limit.

Similar to SN~2023ixf, a possible progenitor has been identified at the location of SN~2024ggi in the images of the Dark Energy Camera Legacy Survey Data Release 10 (DECaLS\footnote{\url{https://www.legacysurvey.org/decamls/}}). We performed photometry for the progenitor in the $g, r, i$, and $z$ band images taken by DECam \citep{DECam-2008H}. Because the $z$-band image has the highest quality, i.e., PSF and signal-to-noise ratio, \texttt{SWarp}\footnote{\url{https://www.astromatic.net/software/swarp/}} \citep{2010ascl.soft10068B} was used to align the astrometric solutions of the $g, r$ and $i$-band images in reference to the $z$-band image. Then we used the dual-image mode of \texttt{SExtractor}\footnote{\url{https://www.astromatic.net/software/sextractor/}} \citep[Source Extractor;][]{Bertin-1996A&AS..117..393B} to do a series of aperture photometry for the progenitor, with aperture diameters from 1.5 to 46.5 pixels for each band image. We chose the aperture with the highest signal-to-noise ratio as the best aperture and performed aperture correction for the corresponding magnitude using the growth curve. A clear detection in the $g,r,i$ and $z$-bands has been noticed, and the estimated $g,r,i$ and $z$ magnitudes are $22.51\pm0.02, 22.74\pm0.02, 22.73\pm 0.03$ and $21.90\pm0.02$, respectively. The quoted errors are the \texttt{SExtractor} estimated uncertainties, which consider only the photon noise and detector noise. The estimated magnitudes are similar to \citet{2024TNSAN.105....1Y} within error limits. It should be noted that the possible progenitor has also been identified in archival images of the Hubble Space Telescope (\texttt{HST}) and VISTA Hemisphere Survey \citep{Srivastav-2024, 2024TNSAN.107....1P}. However, the source is resolved into two sources in the HST F814W band \citep{Srivastav-2024}. 
Therefore, the flux contamination at the SN progenitor location due to the neighboring source may have some influence on the photometric measurements of the progenitor candidate. Consequently, a nonnegligible effect on the estimated magnitudes is possible and should be considered cautiously.

\begin{figure}
\includegraphics[width=8.5cm]{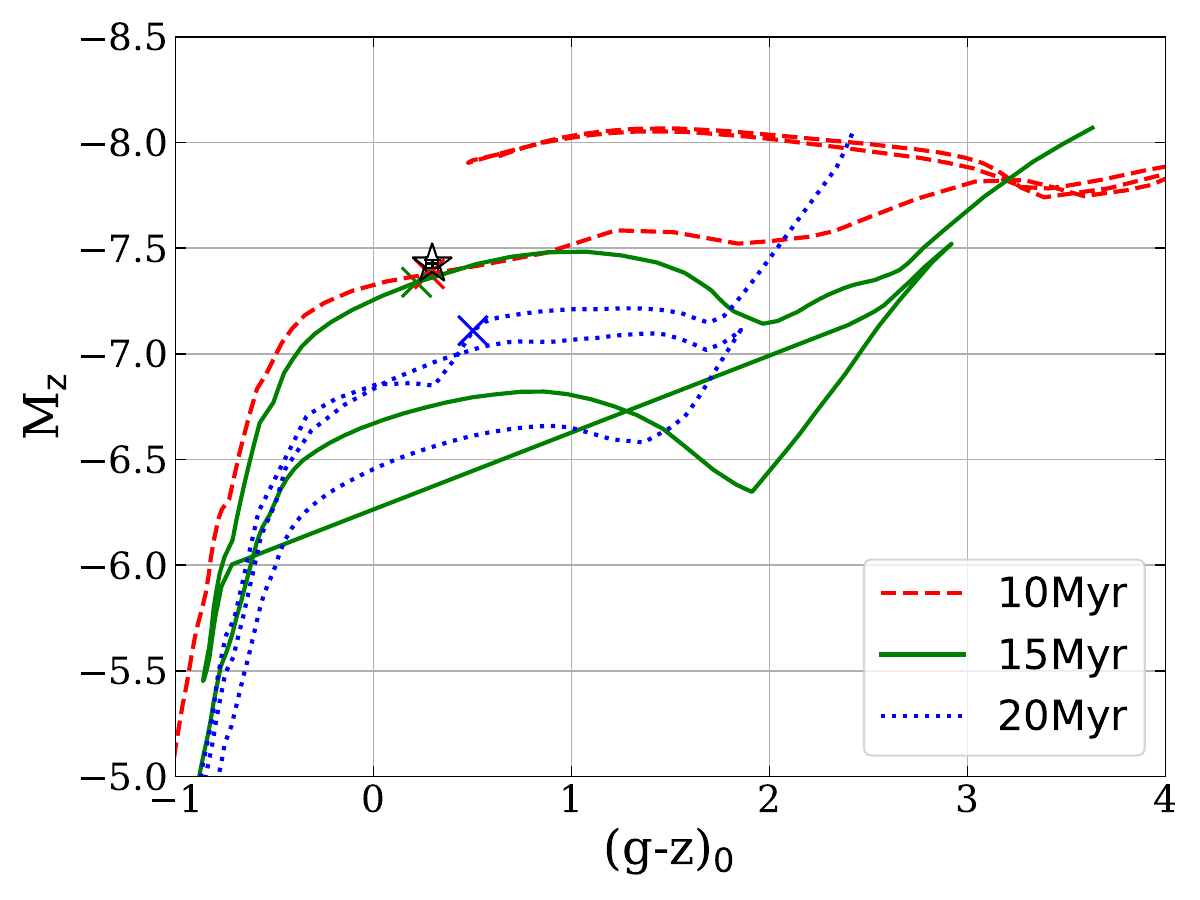}
\caption{The color-magnitude diagram of 10, 15 and 20 Myr \texttt{PARSEC} stellar evolutionary isochrones with metallicity of $Z=0.025$ (red dashed), $Z=0.01$ (green solid), $Z=0.001$ (blue dotted) along with our progenitor candidate. The star symbol is the color-magnitude position measured from DECaLS images. The red, green, and blue cross symbols indicate the initial masses 17, 14, and 11 M$_\odot$, respectively.}
\label{fig:isochrones}
\end{figure}

We used the \texttt{PARSEC}\footnote{\href{http://stev.oapd.inaf.it/cgi-bin/cmd/}{http://stev.oapd.inaf.it/cgi-bin/cmd/}} stellar evolutionary isochrones \citep{2012MNRAS.427..127B} to estimate the mass of the progenitor. The two-part power-law initial mass function is adopted \citep{2001MNRAS.322..231K, 2002Sci...295...82K}. The range of initial mass from 0.1 to 20 $M_\odot$ is explored. We present the progenitor on the color-magnitude diagram of $g$ and $z$ bands (Fig.\,\ref{fig:isochrones}). Again, the same extinction as mentioned in Section~\ref{sec:obs} is used here. The isochrones that match the measurement best have an age of $\sim$10 (red) or 15 (green) Myr, with an initial mass of 17 or 14 $M_\odot$, a metallicity of $Z=0.025$ or $Z=0.01$.
In different studies, the progenitor of SN~2023ixf has been identified as a red supergiant with the estimated mass range between $\sim$9 to 22, e.g., 17 $\pm$ 4 $M_\odot$ \citep{2023ApJ...952L..30J}, 11 $\pm$ 2 $M_\odot$ \citep{Kilpatrick-2023ApJ}, 20--22 $M_\odot$ \citep{Chenxu-2023ApJ}, 8--10 $M_\odot$ \citep{Pledger-2023ApJ}, 20 $\pm$ 4 $M_\odot$ \citep{Soraisam-2023S}, 9--14 $M_\odot$ \citep{Neustadt-2024N}, 14--20 $M_\odot$, \citep{Ransome-2024R}, and 12--14 $M_\odot$ \citep{2023vandyk}. This indicates that the progenitor of SN~2024ggi was a moderately massive star.

Recently, \citet{Xiang-2024arXiv240507699X} investigated the progenitor properties and the local environment of SN~2024ggi using the \texttt{HST} and \texttt{Spitzer} Space Telescope archival images. They derived an initial mass of the possible progenitor $13\pm 1$ $M_\odot$, which is consistent with our estimated lower mass limit (14 $M_\odot$). However, the resolution limitations of DECaLS images should not be overlooked.

The RSG progenitors may be enshrouded in a dusty shell of CSM as revealed in SN~2012aw \citep{Fraser-2012F, VanDyk-2012ApJ...756..131V}, SN~2017eaw \citep{Kilpatrick-2018MNRAS.481.2536K, Rui-2019MNRAS.485.1990R, 2019vandyk}, and SN~2023ixf \citep{Kilpatrick-2023ApJ, Pledger-2023ApJ, Soraisam-2023S, Neustadt-2024N, 2023vandyk}. However, in our investigation of the SN~2024ggi progenitor, the dust mass contribution has not been accounted for, and hence, some enhancement in the estimated mass limit is possible. \citet{Xiang-2024arXiv240507699X} found that the dust shell surrounding the SN~2024ggi progenitor is significantly optically thin (optical depth, $\tau$$_{V}$\,=\,$\sim$0.3) than SN~2023ixf \citep[$\tau$$_{V}$\,=\,$\sim$10--13,][]{Neustadt-2024N, 2023vandyk}. This indicates very limited influence of dust on the SN~2024ggi progenitor's spectral energy distribution. It also implies that possibly there is a much thinner CSM shell around SN~2024ggi progenitor \citep{Xiang-2024arXiv240507699X} in comparison to the dustiest SN~2023ixf progenitor \citep{2023vandyk}.

\section{Summary}\label{summary}

This work presents a good cadence of early light curves (less than a day to about 24 days after explosion) of SN~2024ggi, a nearby Type II SN in the galaxy NGC 3621 ($\sim$6.73 Mpc). Simultaneous multi-band ($ugi$, $vrz$) photometric observations were performed with the 1.6m Multi-channel Photometric Survey Telescope (Mephisto), which is equipped with three channels: Blue ($uv$), Yellow ($gr$) and Red ($iz$). The multi-band color information was used to infer the shock break out, color evolution, and CSM interaction.

The $u-g, g-i, v-r$, and $r−z$ color evolution of SN~2024ggi demonstrate a blue excess in the beginning (two days after first light). Later, $u-g$ and $v-r$ colors display a redward evolution until day $\sim$23 (post-explosion). However, $g-i$ and $r-z$ colors are turning bluer again after $\sim$15 days. SN~2024ggi reached to maximum $M_{g}$=--17.75 mag on day 4 post-explosion, which is towards the bright side in comparison to normal Type II events. The black-body (optical) temperature evolution appears interesting. It increases from 12500 K to 25000 K in 2 days after the explosion and thereafter decreases (Section~\ref{color}). Such characteristics have also been noticed in SN~2018zd and SN~2023ixf. The explosion parameters and progenitor properties were estimated by modeling the light curves with a shock-cooling model (see Section~\ref{LC-properties}). The best-fit progenitor radius of SN~2024ggi is found to be $547^{+14}_{-10}$ R$_\sun$, consistent with a RSG. The comprehensive photometric properties of SN~2024ggi indicate that pure shock-cooling emission may not be attributed to them. Rather, a delayed shock-breakout from a dense CSM is favorable, as also witnessed in SN~2018zd and SN~2023ixf.

We examined the archival images of Dark Energy Camera Legacy Survey Data Release 10 (DECaLS) to search for the possible progenitor of SN~2024ggi. A progenitor candidate was identified in $g,r,i$ and $z$--bands with magnitudes $22.51\pm0.02, 22.74\pm0.02, 22.73\pm 0.03, 21.90\pm0.02$, respectively. However, in HST images, two sources were resolved at the SN location. The best-fitted stellar evolutionary isochrones (solar metallicity) to the extinction-corrected magnitudes indicate a possible progenitor with a mass range of 14\,--\,17 M$_\odot$ (see, Section~\ref{SN-progenitor}). Although a more robust progenitor model can better constrain the initial mass of SN~2024ggi, our estimated mass range (without the dust mass consideration) is towards the lower mass limit of RSGs, as found in the direct identification of Type II SN progenitors. 

It is evident from the observations of nearby SN~2024ggi and SN~2023ixf that prompt and good cadence data are essential to study various aspects of the progenitor, explosion, and surrounding CSM. A coordinated multi-wavelength follow-up is important for investigating the diversity in the light curves, multi-color (temperature) evolution, and implementation of various models. Therefore, the upcoming and existing facilities will play a pivotal role in this regard. The Mephisto facility is expected to be equipped with three mosaic CCD cameras by the end of 2024. The unique feature of simultaneously imaging the particular patch (at each pointing) of the sky in three bands ($ugi$ and $vrz$) will provide a unique opportunity to discover transient sources on the basis of their colors and successive monitoring.

\section{Software and third party data repository citations} \label{sec:soft}

\software{astropy \citep{astropy:2022}, pandas \citep{reback2020pandas}, numpy \citep{harris2020array}, scipy \citep{2020SciPy-NMeth}, Jupyter-notebook \citep{Kluyver2016jupyter}, SWarp \citep{Bertin-2002ASPC..281..228B}, SExtractor \citep{Bertin-1996A&AS..117..393B}, Light Curve Fitting package \citep{2023zndo...8049154H}}.

\section*{Acknowledgments}
We thank the anonymous referee for the insightful suggestions that helped significantly improve the manuscript.
We acknowledge Griffin Hosseinzadeh for the help with the light-curve fitting code. B.K. thanks J. Craig Wheeler and Avinash Singh for important comments and discussions on the manuscript. Mephisto is developed at and operated by the South-Western Institute for Astronomy Research of Yunnan University (SWIFAR-YNU), funded by the ``Yunnan University Development Plan for World-Class University" and ``Yunnan University Development Plan for World-Class Astronomy Discipline". The authors acknowledge support from the ``Science \& Technology Champion Project'' (202005AB160002) and from two ``Team Projects" -- the ``Top Team'' (202305AT350002) and the ``Innovation Team'' (202105AE160021), all funded by the ``Yunnan Revitalization Talent Support Program". 
Y.-Z. Cai is supported by the National Natural Science Foundation of China (NSFC, Grant No. 12303054), the Yunnan Fundamental Research Projects (Grant No. 202401AU070063) and the International Centre of Supernovae, Yunnan Key Laboratory (No. 202302AN360001). We acknowledge the observers and technical support staff for helping with the observations. 

\bibliography{2-SN2024ggi}
\bibliographystyle{aasjournal}

\appendix \label{Apendix}
\begin{figure}
\centering
\includegraphics[width=12cm]{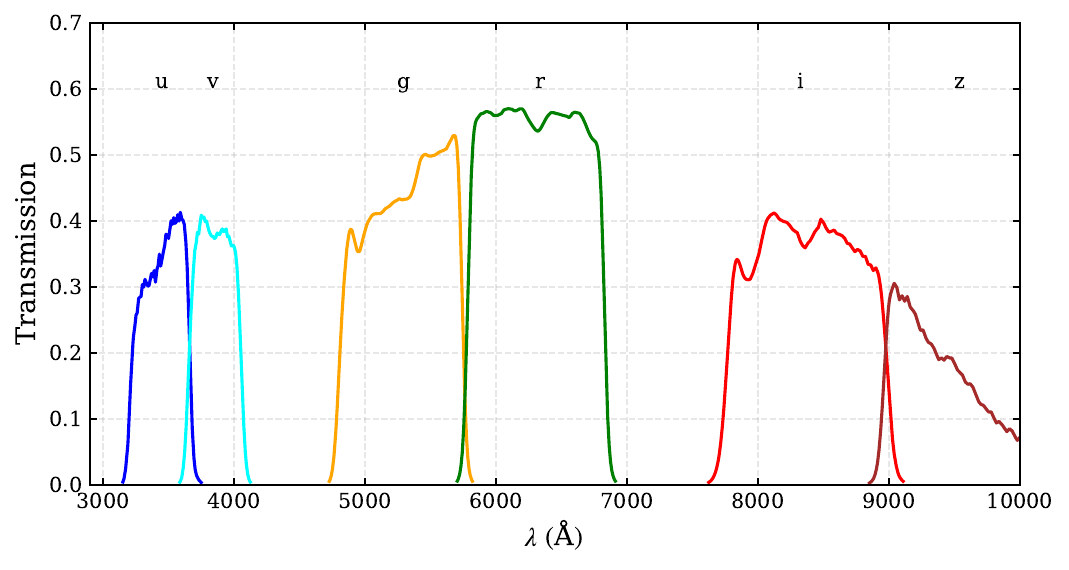}
\caption{The transmission curves of the six filters adopted by the Mephisto telescope.}
\label{fig:filters}
\end{figure}

\section{Mephisto supporting facilities and filter transmission function}
Along with 1.6m Mephisto, there are two 50 cm auxiliary photometric telescopes at the site. The 50 cm array (model: Alluna RC20) has a 505 mm optical aperture and an f/8.1 focal ratio. Both are equipped with an FLI ML 50100 CCD camera of 8176 $\times$ 6132 pixels (of size 6 $\mu$m). The filter systems on these facilities are the same so that coordinated observations can be easily performed (see transients detected with the 1.6m telescope). The $uvgriz$ transmission function is provided in Fig.~\ref{fig:filters}.

\end{document}